\documentclass[final,5p,times,twocolumn]{elsarticle}
\journal{Physics Letters B}
\usepackage{amssymb,amsmath}
\usepackage{slashed}
\allowdisplaybreaks[3]
\usepackage{calrsfs}
\DeclareMathAlphabet{\pazocal}{OMS}{zplm}{m}{n}
\usepackage{hyperref}

\begin{document}

\begin{frontmatter}
\title{
Twist-2 relations for the twist-3 tensor-polarized distribution function 
$f_{LT}$ \\ of a spin-1 hadron by the operator-product-expansion method
}
\author[1,2,3]{S. Kumano}
\ead{kumanos@impcas.ac.cn}
\author[1,2]{Kenshi Kuroki}
\ead{k-kuroki@impcas.ac.cn, corresponding author}
\address[1]{Quark Matter Research Center,
            Institute of Modern Physics, Chinese Academy of Sciences,
            Lanzhou, 730000, China}
\address[2]{Southern Center for Nuclear Science Theory,
            Institute of Modern Physics, Chinese Academy of Sciences,
            Huizhou, 516000, China}
\address[3]{KEK Theory Center,
            Institute of Particle and Nuclear Studies, KEK,
            Oho 1-1, Tsukuba, 305-0801, Japan}

\begin{abstract}
In a spin-1 hadron, tensor-polarized parton distribution functions (PDFs) 
exist. The twist-2 function is $f_{1LL}$ and a twist-3 one is $f_{LT}$. 
Because an experiment is under preparation
at the Thomas Jefferson National Accelerator Facility (JLab)
to measure the cross section of electron-deuteron deep inelastic scattering
with the tensor-polarized deuteron target, 
these PDFs need to be understood theoretically. 
Especially, measurements will be done in a relatively low-$Q^2$ region
at JLab, so that twist-3 contributions could become sizable
in the cross section. In a previous work, a twist-2 relation was derived 
for $f_{LT}$ in terms of $f_{1LL}$ by using a nonlocal operator, 
and it corresponds to the Wandzura-Wilczek (WW) relation between 
$g_1$ and $g_2$. In addition, another relation similar to 
the Burkhardt-Cottingham (BC) sum rule was obtained. 
It is known that a formal way to derive the WW relation and the BC sum rule
is to use the operator product expansion (OPE) with local operators.
In this work, the WW-like relation and the BC-like sum rule for $f_{LT}$ 
are derived by using the local OPE method as a reliable independent way 
to establish these relations.
\end{abstract}


\end{frontmatter}

\date{July 7, 2026}

\section{Introduction}
\label{introduction}

Spin structure of the spin-1/2 nucleon has been investigated in details
at high energies for finding the origin of nucleon spin.
Hadrons with higher spin $\ge 1$ 
have much richer spin structure because of the existence 
of tensor polarizations. There are recent theoretical developments on 
polarized parton distribution functions (PDFs) of 
spin-1 \cite{Kumano:2024fpr} and spin-3/2 \cite{Fu:2026mb} hadrons.
This additional spin structure has not been investigated extensively
in comparison with the one of the spin-1/2 nucleon
partly because of a lack of an experimental project for a long time. 
However, time has come to investigate this topic seriously
because an electron deep inelastic scattering (DIS) experiment
is under preparation now at the Thomas Jefferson National 
Accelerator Facility (JLab) with the tensor-polarized deuteron
\cite{jlab-b1-2023,Poudel:2025nof}.
There are other experimental facilities where tensor-polarized
deuteron experiments will be possible in several years, 
such as the Fermi National Accelerator Laboratory (Fermilab)
\cite{Fermilab-SeaQuest-2017,Keller:2020wan}, 
the Nuclotron-based Ion Collider fAcility (NICA) \cite{Arbuzov:2020cqg}, 
the Large Hadron Collider (LHC) \cite{Aidala:2019pit},
and Electron-Ion Colliders (EICs) \cite{AbdulKhalek:2021gbh,Anderle:2021wcy}.

In order to understand such upcoming experimental measurements
on the spin-1 deuteron, it is necessary to investigate 
the tensor-polarized structure functions and PDFs theoretically.
In particular, JLab data are expected to contain significant
higher-twist effects because they are obtained typically 
in a few GeV$^2$ $Q^2$ region.
In addition to the twist-2 functions \cite{Bacchetta:2000jk},
possible twist-3 and 4 PDFs and transverse-momentum-dependent parton
distribution functions (TMDs) were obtained for spin-1 hadrons
in Ref.\,\cite{Kumano:2020ijt}.
Among them, there are the twist-2 PDF $f_{1LL}$ and the twist-3 one $f_{LT}$.
It was shown in Ref.\,\cite{Kumano:2021fem} that there exists
a twist-2 relation between $f_{1LL}$ and $f_{LT}$
in the similar way to the Wandzura-Wilczek (WW) relation
\cite{Wandzura:1977qf} between the polarized structure functions 
$g_1$ and $g_2$. It was also shown that the deviation from 
this WW-like relation indicates twist-3 multiparton distribution functions,
and this deviation part is called the dynamical (or genuine) twist-3 term.
In addition, a sum rule which is similar to 
the Burkhardt-Cottingham (BC) sum rule \cite{Burkhardt:1970ti}
was also obtained \cite{Kumano:2021fem}.
Based on these twist-2 relations, the function $f_{LT}$ of the deuteron
was numerically estimated for the first time \cite{Kumano:2025rai}.

In the derivation of the BC-like sum rule and WW-like relation 
\cite{Kumano:2021fem}, a nonlocal operator was used 
with lightcone correlation functions,
whereas a formal way to show such relations
is to use the operator product expansion (OPE) with local operators.
In addition, it is known that the original WW relation is
related to the rotational invariance 
\cite{Jaffe:1989xx,Jaffe:1996zw,Anikin:2001ge}.
It is desirable to show the WW-like relation
and the BC-like sum rule by the traditional local OPE method, in which
the rotational invariance is manifestly satisfied,
as a reliable independent method to confirm the relations
for establishing them.

In this work, the WW-like relation and the BC-like sum rule are derived
for $f_{LT}$ by using the local OPE method.
This article consists of the following.
In Section\,\ref{spin-1-pdfs}, the tensor-polarized PDFs $f_{1LL}$ and $f_{LT}$
are introduced for spin-1 hadrons. 
The WW-like relation and BC-like sum rule are derived 
by the local OPE method in Section\,\ref{ope-spin-1},
and our studies are summarized in Section~\ref{summary}.

\section{Tensor-polarized parton distribution functions\\ in a spin-1 hadron}
\label{spin-1-pdfs}

The PDFs of a hadron are generally defined in the correlation function 
with a nonlocal operator as
\begin{align}
\Phi_{ij}^{[c]} (k, P, T) = \! \! \int \! \frac{d^4 \xi}{(2\pi)^4} \, 
   e^{ i k \cdot \xi} \langle  P, T \left | \, \bar\psi _j (0) 
   W^{[c]} (0, \xi) \psi _i (\xi)  \, \right | P, T  \rangle ,
\nonumber\\[-0.20cm]
\label{eqn:correlation-q}
\end{align} 
\vspace{-0.70cm}

\noindent
where $\psi$ is the quark field, $P$ and $T$ indicate the hadron
momentum and tensor polarization, $i$ and $j$ are components of $\psi$,
$k$ is the quark momentum, and $W^{[c]}$ is the gauge link 
for the color gauge invariance with the integral path $c$.
The correlation function is the amplitude to extract a parton 
from the hadron and to insert it at a different space-time point $\xi$.
We do not write the spin vector polarization $S$ explicitly 
in Eq.\,(\ref{eqn:correlation-q}) because only the tensor polarization $T$, 
which is specific to hadrons with spin$\ge 1$, is investigated in this paper. 
The vector polarization part is essentially the same as the one 
for the spin-1/2 nucleons.
The collinear correlation function is obtained by integrating
Eq.\,(\ref{eqn:correlation-q}) over the parton momentum as
\begin{align}
& \hspace{-0.30cm}
\Phi_{ij} (x, P, T ) 
    = \int d^2 k_T \, dk^+ dk^- \, \Phi^{[c]}_{ij} (k, P, T ) \, 
    \delta (k^+  -x P^+) 
\nonumber \\
& \hspace{-0.51cm}
    = \! \int \frac{d\xi^{\, -}}{2\pi} \, e^{i x P^+ \xi^{\, -}} 
    \langle \, P , T \left | \, \bar\psi _j (0)  
    \,  W (0, \xi \, | \, n)   \psi _i (\xi)  \, \right | 
       P, \,  T \, \rangle _{\xi^+ =0, \, \vec\xi_T=0} ,
\label{eqn:correlation-pdf}
\end{align}
where the lightcone $\pm$ momenta are defined by 
$a^\pm = (a^0 \pm a^3)/\sqrt{2}$, and
$x$ is the lightcone momentum fraction given by $k^+ = x P^+$.
The gauge link $W (0,\xi \, | \, n)$ indicates the line connecting 
$0$ to $\xi = (\xi^{\,+}, \xi^{\,-}, \vec \xi_T)$ 
along the straight lightcone direction of $\xi^{\,-}$.
The lightcone vectors $n^{\,\mu}$ and $\bar n^{\,\mu}$ are
defined by
\begin{align}
    n^{\,\mu} =\frac{1}{\sqrt{2}} (\, 1,\, 0,\, 0,\,  -1 \, ), \ \ \
    \bar n^{\,\mu} =\frac{1}{\sqrt{2}} (\, 1,\, 0,\, 0,\,  1 \, ) .
\label{eqn:lightcone-n-nbar}
\end{align}

The tensor polarization $T^{\mu\nu}$ is expressed 
in terms of the tensor-polarization parameters 
$S_{LL}$, $S_{LT}^{\mu}$, and $S_{TT}^{\mu\nu}$ as
\cite{Bacchetta:2000jk,Kumano:2020ijt}
\footnote{One may note that the symmetrization 
   $a^{\{ \mu} b^{\nu \}} = a^\mu b^\nu + b^\mu a^\nu$
   is used in Refs.\,\cite{Kumano:2020ijt,Kumano:2021fem}
   in expressing $T^{\mu\nu}$. However, the symmetrization convention 
   $a^{\{ \mu} b^{\nu \}} = (a^\mu b^\nu + b^\mu a^\nu)/2!$
   is often used in OPE studies as defined below
   Eq.\,(\ref{eqn:vector-matrix-LHS-2}) with the factor $1/n!$,
   so that the index symmetrization $\{ \cdots \}$
   is not used in this paper for defining $T^{\mu\nu}$ 
   to avoid the confusion.}
\begin{align}
\hspace{-0.40cm}
T^{\mu\nu} = \frac{1}{2} & \left [\frac{4}{3} S_{LL} \frac{(P\cdot n)^2}{M^2} 
               \bar n^{\,\mu} \bar n^{\,\nu} 
          - \frac{2}{3} S_{LL} \left( \bar n^{\,\mu} n^{\nu} 
                  + \bar n^{\nu} n^{\,\mu} -g_T^{\,\mu\nu} \right)
\right.
\nonumber \\
& \hspace{-0.0cm}
+ \frac{1}{3} S_{LL} \frac{M^2}{(P\cdot n)^2}n^{\,\mu} n^{\,\nu}
+ \frac{P \cdot n}{M}
  \left( \bar n^{\,\mu} S_{LT}^{\nu} + \bar n^{\nu} S_{LT}^{\mu}  \right)
\nonumber \\
& \hspace{-0.0cm}
\left.
- \frac{M}{2 P \cdot n} \left( n^{\,\mu} S_{LT}^{\nu} +n^{\nu} S_{LT}^{\mu} \right)
+ S_{TT}^{\mu\nu} \right ].
\label{eqn:spin-1-tensor-1}
\end{align}
where $g_T^{\,\mu\nu}$ is given by 
$g_T^{\,\mu\nu} = g^{\,\mu\nu} 
 - (\bar n^{\,\mu} n^{\nu}+ \bar n^{\nu} n^{\,\mu} ) $
and $M$ is the hadron mass.
The tensor $T^{\mu\nu}$ is symmetric and traceless, 
and it satisfies $P_\mu T^{\mu\nu}=0$.
The parameter $S_{LL}$ is the tensor polarization along the longitudinal axis,
and $S_{LT}^{\mu}$ and $S_{TT}^{\mu\nu}$ indicate 
polarization differences along the axes 
between the longitudinal and transverse directions 
and along the transverse axes, respectively~\cite{Bacchetta:2000jk,Kumano:2020ijt}.
They satisfy the relations 
$S_{LT} \cdot \bar n = S_{LT} \cdot n =0$ and
$S_{TT}^{\mu\nu} \,\bar n_{\mu}=S_{TT}^{\mu\nu} \, n_{\mu}
 = S_{TT}^{\mu\nu} g_{T \mu\nu} =0$.
The collinear correlation function is expressed 
in terms of these parameters and tensor-polarized PDFs
$f_{1LL} (x)$, $e_{LL} (x)$, $f_{LT} (x)$, and $f_{3LL} (x)$ as
\cite{Kumano:2021fem}
\begin{align}
\Phi (x,P,T) 
    = & \frac{1}{2} \, \bigg[ \, S_{LL} \, \slashed{\bar n} \,
     f_{1LL} (x) + \frac{M}{P \cdot n} \, S_{LL} \, e_{LL} (x) 
\nonumber \\
    & + \frac{M}{P \cdot n} \, \slashed{S}_{LT} \, f_{LT} (x) 
    + \frac{M^2}{(P \cdot n)^2} \, S_{LL} \, \slashed{n} \,
     f_{3LL} (x) \, \bigg] ,
\label{eqn:collinear-correlation-pdfs}
\end{align}
where $\slashed{a} = \gamma^{\,\mu} a_\mu$.
The quark distribution function $f_{1LL} (x)$ is a twist-2 function, 
while $e_{LL} (x)$ and $ f_{LT} (x)$ are twist-3 functions, 
and $f_{3LL} (x)$ is a twist-4 function. 
The $e_{LL} (x)$ is a chiral-odd function, and the others
are chiral-even ones.

In the previous derivation of the WW-like relation and the BC-like sum rule
in 2021 \cite{Kumano:2021fem},
the nonlocal operator $\bar\psi (0) \gamma^{\,\mu} \psi (\xi)$ was used. 
Its matrix element was expressed by
the available Lorentz vectors $P^{\,\mu}$, $\xi^{\,\mu}$,
and $T^{\mu\nu} \xi_\nu$ as 
\begin{align}
& \ \hspace{-0.5cm} \langle \, P ,  T \left | \,  \bar\psi(0)  \gamma^{\,\mu}  \psi(\xi) 
  \, \right | P \,  , T \, \rangle
\nonumber \\ 
& \ \hspace{-0.5cm}
= \! \int_{-1}^1 \! dx \, e^{-i x P\cdot \xi}
  \left[ \,  \xi \cdot T \cdot \xi  \left \{  A(x) \, P^{\,\mu}
          + B(x) \, \xi^{\,\mu} \right \}
          + C(x) \, T^{\mu\nu} \xi_\nu
  \, \right]  .
\end{align}
The coefficients $A$, $B$, and $C$ can be obtained
in arbitrary frame, so that they were determined on the lightcone.
Then, the twist-3 combination 
$\bar \psi (0) ( \gamma^{\,\mu} \partial^{\nu} 
  - \gamma^{\nu} \partial^{\,\mu} ) \, \psi (\xi)$
was used for finding the WW-like relation and also
the twist-3 multi-parton distribution functions, which
are deviation terms from the WW-like relation.
Then, the BC-like sum rule was obtained.
In Section\,\ref{ope-spin-1}, we derive 
the WW-like twist-2 relation and the BC-like sum rule
by a formal OPE method with local operators
as an independent way.

\section{Operator-product-expansion method for deriving\\ the twist-2 relations}
\label{ope-spin-1}

The OPE is used for deriving the twist-2 relations in this section.
The OPE method for deriving the WW relation 
and BC sum rule is summarized in
Refs.\,\cite{Jaffe:1989xx,Jaffe:1996zw,Kodaira:1998jn}.
The polarized structure function $g_2$ of the spin-1/2 nucleon
is separated into WW and dynamical twist-3 parts as
\begin{align}
g_2 (x) = g_2^{\text{WW}} (x) & +  g_2^{\text{(HT)}} (x), 
\label{eqn:g2-twist-2-3}
\end{align} 
where HT indicates higher twist,
and then the WW part is expressed by the twist-2 
structure function $g_1$ as
\begin{align}
g_2^{\text{WW}} (x) & = - g_1 (x) + \int_x^1 \frac{dy}{y} g_1 (y) .
\label{eqn:ww}
\end{align} 
This is the WW relation.
If this WW relation is integrated over $x$, 
it becomes
\begin{align} 
\int_0^1 dx \, g_2^{WW} (x) = 0 ,
\label{eqn:bc-sum}
\end{align}
which is the BC sum rule.

In the following derivations, we follow the explanations
in Ref.\,\cite{Kodaira:1998jn} on the OPE method
for deriving the WW relation and the BC sum rule
to obtain the corresponding twist-2 relations for $f_{LT}$.
The tensor-polarization part of the matrix element 
of the nonlocal vector operator 
$ \bar\psi (0) \, \gamma^{\,\sigma} \, \psi (\xi)$ is given by
\cite{Kumano:2021fem}
\begin{align}
& \hspace{-0.00cm}
\left.
\int \frac{d(P \cdot \xi)}{4\pi P \cdot n} e^{\,ixP \cdot \xi}
\langle \, P , T \left | \, \bar\psi (0) 
\, \gamma^{\,\sigma} \, 
\psi (\xi)  \, \right | \! P,  T \,
\rangle \, \right| _{\,\xi^+ =0, \, \vec\xi_T=0} 
\nonumber \\
& \hspace{0.00cm}
= 
S_{LL} \bar n^{\,\sigma} f_{1LL} (x) 
+  \frac{M}{P\cdot n}  S_{LT}^{\,\sigma} f_{LT} (x) ,
\label{eqn:vector-matrix-1}
\end{align}
in terms of the tensor-polarized PDFs 
$f_{1LL}$ and $f_{LT}$.
The twist-4 distribution $f_{3LL}$ is neglected
in this work up to twist 3.
Taking $n$-th moment of Eq.\,(\ref{eqn:vector-matrix-1}), 
we obtain
\begin{align}
& 
\left.
\int_{-1}^1 dx \, x^{\, n-1} \! \!
\int \frac{d(P \cdot \xi)}{4\pi P \cdot n} e^{\, ixP \cdot \xi}
\langle \, P , T \left | \, \bar\psi (0) 
\, \gamma^{\,\sigma} \, 
\psi (\xi)  \, \right | \! P, T \,
\rangle \, \right| _{\,\xi^+ =0, \, \vec\xi_T=0} 
\nonumber \\
& \hspace{0.00cm}
= 
S_{LL} \bar n^{\,\sigma}  \int_{-1}^1 dx \, x^{\, n-1} f_{1LL} (x) 
+ \frac{M}{P\cdot n}  S_{LT}^{\,\sigma} 
     \int_{-1}^1 dx \, x^{\, n-1} f_{LT} (x) .
\label{eqn:vector-matrix-2}
\nonumber\\[-0.20cm]
\end{align}

\vspace{-0.20cm}
The quark field $\psi (\xi)$ is expanded in 
a Taylor series as
\begin{align}
\psi (\xi) =  
\sum_{m=0}^\infty  \frac{1}{m!}
    (\xi \cdot \partial )^m \psi (\xi)_{\xi=0} .
\label{eqn:psi-expansion}
\end{align}
The factor $x^{\, n-1}$ on the left-hand side of 
Eq.\,(\ref{eqn:vector-matrix-2})
is obtained by taking derivatives of
$e^{\, ixP \cdot \xi}$ with respect to $P \cdot \xi$.
Then, the left-hand side of Eq.\,(\ref{eqn:vector-matrix-2}) becomes
\begin{align}
& \! \! \!
\left.
\int_{-1}^1 dx \, x^{\, n-1}
\int \frac{d(P \cdot \xi)}{4\pi P \cdot n} e^{\, ixP \cdot \xi}
\langle \, P , T \left | \, \bar\psi (0) 
\, \gamma^{\,\sigma} \, 
\psi (\xi)  \, \right | \! P, T \,
\rangle \, 
\right|_{\,\xi^+ =0, \, \vec\xi_T=0} 
\nonumber \\
& \! \! \!
= \int_{-1}^1 dx 
\int \frac{d(P \cdot \xi)}{4\pi P \cdot n} \frac{1}{i^{\, n-1}}
\left [ \frac{\partial^{\, n-1}}{\partial (P \cdot \xi)^{\, n-1}}
 e^{\, ixP \cdot \xi} \right ]
\nonumber \\
& \hspace{-0.14cm}
\left.
\times
\left\langle \, P , T \left | \, \bar\psi (0) 
\, \gamma^{\, \sigma} \, 
 \sum_{m=0}^\infty  \frac{1}{m!}
    (\xi \cdot \partial )^m \psi (\xi)_{\xi=0} 
  \, \right | \! P,  T \,
\right\rangle \, \right| _{\,\xi^+ =0, \, \vec\xi_T=0} .
\label{eqn:vector-matrix-LHS-1}
\end{align}
Shifting the derivative operation to the matrix-element part
by the partial integration 
and noting the relation
\begin{align}
& \partial^{\,\mu_1} \cdots \partial^{\,\mu_{n-1}} \!
     \sum_{m=0}^\infty  \frac{1}{m!} (\xi \cdot \partial )^m 
=  \partial^{\, \{ \, \mu_1} \cdots \partial^{\, \mu_{n-1} \}}
\left ( 1 + \sum_{m=1}^\infty  \frac{1}{m!} (\xi \cdot \partial )^m \right ) ,
\label{eqn:derivative-part}
\end{align}
\\[-0.60cm]
we obtain 
\begin{align}
& \! \! \! \!
\left.
\int_{-1}^1 dx \, x^{\, n-1} \! \!
\int \frac{d(P \cdot \xi)}{4\pi P \cdot n} e^{\, ixP \cdot \xi}
\langle \, P , T \left | \, \bar\psi (0) 
\, \gamma^{\,\sigma} \, 
\psi (\xi)  \, \right | \! P, \,  T \,
\rangle \, \right| _{\,\xi^+ =0, \, \vec\xi_T=0} 
\nonumber \\
& \hspace{-0.41cm}
= \frac{n_{\mu_1} \cdots n_{\mu_{n-1}}}{2 \, (P \cdot n)^{\, n}} \!
\left\langle \, P , T \left | \, i^{\, n-1} \bar\psi (0) 
\, \gamma^{\,\sigma} \, 
    \partial^{\, \{\,\mu_1} \cdots \partial^{\,\mu_{n-1} \}}  \psi (0)
  \, \right | \! P,  T \,
\right\rangle ,
\label{eqn:vector-matrix-LHS-2}
\end{align}
for $n\ge 2$.
The curly bracket $\{ \cdots \}$ indicates the symmetrization of indices
defined by 
$\partial^{\, \{\,\mu_1} \cdots \partial^{\,\mu_{n-1} \}}
= ( \, \partial^{\,\mu_1} \cdots \partial^{\,\mu_{n-1}} 
   + \text{permutations} \, ) \, / \, (n-1)!$.
The derivatives are replaced by the covariant ones by
the minimal substitution, then the local operator becomes
\begin{align}
R^{\, \sigma \{ \mu_1 \cdots \mu_{n-1} \}}
 = i^{\, n-1} \bar\psi (0) \gamma^{\,\sigma} 
      D^{\, \{ \mu_1} \cdots D^{\, \mu_{n-1} \} } \psi (0) 
 - \text{traces}.
\label{eqn:operator-R}
\end{align} 
Here, the traces indicate that the higher-twist ($\ge 4$) trace terms 
$g_{\mu_i \mu_j} R^{\sigma \{ \mu_1 \cdots \mu_{n-1} \}}$
and $g_{\sigma \mu_i} R^{\sigma \{ \mu_1 \cdots \mu_{n-1} \}}$
are subtracted to make the operator traceless,
and it becomes the operator up to twist 3.
Therefore, the twist-4 PDF $f_{3LL} (x)$ does not appear in this work.
From Eqs.\,(\ref{eqn:vector-matrix-2}), (\ref{eqn:vector-matrix-LHS-2}),
and (\ref{eqn:operator-R}), we obtain
\begin{align}
& \hspace{-0.20cm}
\frac{n_{\,\mu_1} \cdots n_{\,\mu_{n-1}}}{2 \, (P \cdot n)^{\, n}} 
\left\langle \, P , T \left | \, 
R^{\, \sigma \{ \mu_1 \cdots \mu_{n-1} \}} 
\, \right | \! P,  T \, \right\rangle 
\nonumber \\
& \hspace{-0.30cm}
= 
S_{LL} \bar n^{\, \sigma} 
\! \int_{-1}^1 \! dx \, x^{\, n-1} f_{1LL} (x) 
+ \frac{M}{P\cdot n}  S_{LT}^{\, \sigma} 
    \! \int_{-1}^1  \! dx \, x^{\, n-1} f_{LT} (x) .
\label{eqn:vector-matrix-LHS-3}
\end{align}
Then, the operator is decomposed into symmetric
and mixed symmetric terms as
\begin{align}
R^{\, \sigma \{ \mu_1 \cdots \mu_{n-1} \}}
& = R^{\, \{ \sigma \mu_1 \cdots \mu_{n-1} \}}
  + R^{\, [ \sigma \{ \mu_1 ] \cdots \mu_{n-1} \}} ,
\nonumber \\
R^{\, \{ \sigma \mu_1 \cdots \mu_{n-1} \}}
& = \frac{1}{n}
\left [  
  R^{\, \sigma \{ \mu_1 \mu_2 \cdots \mu_{n-1} \}}
+ R^{\, \mu_1  \{ \sigma \mu_2 \cdots \mu_{n-1} \}}
+ R^{\, \mu_2  \{ \mu_1 \sigma \cdots \mu_{n-1} \}}
+ \cdots
\right ] ,
\nonumber \\
R^{\, [ \sigma \{ \mu_1 ] \cdots \mu_{n-1} \}}
& = \frac{1}{n}
\left [  
(n-1) R^{\, \sigma \{ \mu_1 \mu_2 \cdots \mu_{n-1} \}}
- R^{\, \mu_1  \{ \sigma \mu_2 \cdots \mu_{n-1} \}}   \right.
\nonumber \\
& \hspace{0.70cm}  \left.
- R^{\, \mu_2  \{ \mu_1 \sigma \cdots \mu_{n-1} \}}
- \cdots
\right ] ,
\label{eqn:twist-2-3}
\end{align} 
where the square bracket $[ \cdots ]$ indicates antisymmetric combination.
One may note that the antisymmetric combinations are taken between $\sigma$
and all the other indices $\mu_1$, $\cdots$, $\mu_{n-1}$ 
on the right-hand side of  Eq.\,(\ref{eqn:twist-2-3}),
although the square bracket is written only between $\sigma$ and $\mu_1$ 
on the left hand side.
The twist is defined by the operator mass dimension minus its spin.
The symmetric operator $R^{\, \{ \sigma \mu_1 \cdots \mu_{n-1} \}}$
and the mixed symmetric one $R^{\, [ \sigma \{ \mu_1 ] \cdots \mu_{n-1} \}}$
are twist-2 and 3 operators, respectively.

In general, these matrix elements of the local operators should be
expressed in terms of the available Lorentz vector $P^{\,\mu}$ and 
the tensor $T^{\mu\nu}$ as
\begin{align}
& \! \! \!
\langle \, P, T \, \big | \, 
R^{\, \{ \sigma \mu_1 \cdots \mu_{n-1} \}} \, 
\big | \, P, T \, \rangle 
= \frac{2}{n} a_n M^2
\left [ \,
\sum_{i=1}^{n-1} \left ( T^{\sigma \mu_i} + T^{\mu_i \sigma} \right )
\! \prod_{j (\ne i) =1}^{n-1} P^{\, \mu_j}
\right.
\nonumber \\
& \hspace{3.0cm} \left.
+ \sum_{i=1}^{n-1} \sum_{j (\ne i) =1}^{n-1} 
   T^{\mu_i \mu_j} P^{\,\sigma} 
\! \prod_{k (\ne i,j) =1}^{n-1} P^{\, \mu_k} \, \right] ,
\\
& \! \! \!
\langle \, P, T \, \big | \, 
R^{\, [ \sigma \{ \mu_1 ] \cdots \mu_{n-1} \}} \, 
\big | \, P, T \, \rangle 
= \frac{2}{n} d_n M^2
\nonumber \\
& \hspace{1.5cm}
\times
\sum_{i=1}^{n-1} \sum_{j (\ne i) =1}^{n-1} 
 \left( P^{\,\sigma} T^{\mu_i \mu_j} -  P^{\, \mu_i} T^{\sigma \mu_j}  \right)
\prod_{k (\ne i,j)=1}^{n-1} P^{\,\mu_k} ,
\label{eqn:matrix-twist-2-3}
\end{align} 
where $a_n$ and $d_n$ are overall constants.
In order to calculate the left-hand side of 
Eq.\,(\ref{eqn:vector-matrix-LHS-3}) by 
using Eq.\,(\ref{eqn:matrix-twist-2-3}),
we first need to calculate $n_{\,\mu} T^{\,\mu\nu}$ and 
$n_{\,\mu} n_{\,\nu} T^{\,\mu\nu}$.
From Eq.\,(\ref{eqn:spin-1-tensor-1}) and (\ref{eqn:lightcone-n-nbar}),
they are obtained as
\begin{align}
n_{\,\mu} T^{\,\mu\nu} 
    & = \frac{2}{3} \, S_{LL} \, \frac{(P\cdot n)^2}{M^2} \, \bar n^{\,\nu}
      + \frac{P \cdot n}{2M} \, S_{LT}^{\, \nu} 
      - \frac{1}{3} \, S_{LL} \, n^{\,\nu} , 
\nonumber \\
n_{\,\mu} n_{\,\nu} T^{\,\mu\nu} 
    &  = \frac{2}{3} \, S_{LL} \, \frac{(P\cdot n)^2}{M^2} .
\label{eqn:ndotTmunu}
\end{align} 
In addition, the hadron momentum is expressed as
\cite{Kumano:2020ijt}
\begin{align}
P^{\,\mu} = (P \cdot n) \, \bar n^{\,\mu} 
           + \frac{M^2}{2 P \cdot n} \, n^{\,\mu} .
\label{eqn:hadron-momentum}
\end{align} 
By using these relations, the left-hand side of 
Eq.\,(\ref{eqn:vector-matrix-LHS-3}) becomes
\begin{align}
& \hspace{-0.00cm}
\frac{n_{\, \mu_1} \cdots n_{\, \mu_{n-1}}}{2 \, (P \cdot n)^{\, n}} 
\left\langle \, P , T \left | \, 
R^{\, \sigma \{ \mu_1 \cdots \mu_{n-1} \}} 
\, \right | \! P,  T \, \right\rangle 
\nonumber \\
& \hspace{-0.10cm}
= 
S_{LL} \bar n^{\,\sigma}  \left[ \frac{2(n-1)}{3} a_n \right]
+ \frac{M}{P\cdot n}  S_{LT}^{\, \sigma} 
      \left[ \frac{n-1}{n} a_n - \frac{(n-1)(n-2)}{2n} d_n \right] ,
\label{eqn:vector-matrix-LHS-4}
\end{align}
\\[-0.50cm]
where the twist-4 term $O(M^2/(P\cdot n)^2)$ is neglected.

The antiquark distribution is given by the distribution
at negative $x$ as $\bar f (x) = - f(-x)$ 
for the chiral-even functions $f_{1LL}(x)$ and $f_{LT}(x)$
\cite{Kumano:2021fem}
and the distributions $f ^{\,\pm} (x)$ are defined by
\begin{align}
f ^{\,\pm} (x) \equiv f (x) \pm \bar f(x) .
\label{eqn:f+-}
\end{align}
Identifying each term of the left-hand sides
of Eqs.\,(\ref{eqn:vector-matrix-LHS-3}) and
(\ref{eqn:vector-matrix-LHS-4}), we obtain 
the moments of the tensor-polarized PDFs as
\begin{align}
\int_{-1}^1 dx \, x^{\, n-1} f_{1LL} (x) 
& = \int_0^1 dx \, x^{\, n-1} f_{1LL}^{\,\pm} (x) 
= \frac{2(n-1)}{3} a_n ,
\nonumber \\
\int_{-1}^1 dx \, x^{\, n-1} f_{LT} (x) 
& 
= \int_0^1 dx \, x^{\, n-1} f_{LT}^{\,\pm} (x) 
\nonumber \\
&
= \frac{n-1}{n} a_n - \frac{(n-1)(n-2)}{2n} d_n ,
\label{eqn:f1LL-f1LT-moments}
\end{align} 
where $f^{\,+}$ is for even integer $n=2,\,4,\,\cdots$
and $f^{\,-}$ is for odd integer $n=1,\,3,\,\cdots$.

For the distribution $f(x)$ defined in the range $0 \le x \le 1$,
we have the following relation
\begin{align}
\int_0^1 dx \, x^{\, n-1} \int_x^1 \frac{dy}{y} f(x) 
= \frac{1}{n} \int_0^1 dx \, x^{\, n-1} f(x) .
\label{eqn:dxdyfx}
\end{align} 
From the moments of $f_{1LL}^{\,\pm}$ and $f_{LT}^{\,\pm}$
in Eq.\,(\ref{eqn:f1LL-f1LT-moments}), 
one obtains by using Eq.\,(\ref{eqn:dxdyfx})
\begin{align}
\hspace{-0.25cm}
\int_0^1 dx \, x^{\, n-1} f_{LT}^{\, \pm} (x) 
& = 
\int_0^1 dx \, x^{\, n-1}
\left[
\frac{3}{2} \int_x^1 \frac{dy}{y} f_{1LL}^{\, \pm} (y) \right]
\nonumber\\
& \hspace{0.30cm}
- \frac{(n-1)(n-2)}{2n} d_n .
\label{eqn:f1LL-f1LT-1}
\end{align} 
On the right-hand side, the first term comes from
the twist-2 term and the second one is from
the dynamical twist-3 term as given by the coefficient $d_n$.
This equation indicates that 
the PDF $f_{LT} (x)$ is written in terms of
the WW-like term $f_{LT}^{\, \pm \,\text{twist-2}} (x)$
and the remaining dynamical twist-3 one $f_{LT}^{\,\pm\,\text{(HT)}} (x)$
\begin{align}
&  f_{LT}^{\,\pm} (x) 
= f_{LT}^{\, \pm \,\text{twist-2}} (x)  + f_{LT}^{\,\pm\,\text{(HT)}} (x),
\label{eqn:fltpm}
\\
& \hspace{1.0cm} 
f_{LT}^{\, \pm \,\text{twist-2}} (x) 
= \frac{3}{2} \int_x^1 \frac{dy}{y} f_{1LL}^{\, \pm} (y) , 
\label{eqn:fltpm-twist2}
\\
& \hspace{1.0cm} 
\int_0^1 dx \, x^{\, n-1} f_{LT}^{\,\pm\,\text{(HT)}} (x) 
= - \frac{(n-1)(n-2)}{2n} d_n .
\label{eqn:fltpm-twist3}
\end{align}
Here, $f_{LT}^{\, \pm \,\text{twist-2}}$ is
the twist-3 function part determined by the twist-2 function $f_{1LL}$,
and it contains kinematical and intrinsic terms
\cite{Kanazawa:2015ajw,Song:2023ooi}.
Equation (\ref{eqn:fltpm-twist2}) is the WW-like relation 
obtained in Ref.\,\cite{Kumano:2021fem}.
If we define the distribution $f_{2LT}^{\,\pm}(x)$ by
\begin{align}
f_{2LT}^{\, \pm \,\text{twist-2}} (x) = \frac{2}{3} f_{LT} ^{\,\pm}(x) 
    - f_{1LL} ^{\,\pm}(x) ,
\label{eqn:f2LT}
\end{align} 
Eq.\,(\ref{eqn:fltpm-twist2}) becomes
\begin{align}
f_{2LT}^{\, \pm \,\text{twist-2}} (x) 
= - f_{1LL}^\pm (x) 
+ \int_x^1 \frac{dy}{y} f_{1LL}^\pm (y) ,
\label{eqn:WW-like-2}
\end{align}
which is the same form as the original WW relation
in Eq.\,(\ref{eqn:ww}). Integrating this equation over $x$, 
we obtain
\begin{align}
\int_0^1 dx \, f_{2LT}^{\, \pm \,\text{twist-2}} (x) =0 ,
\label{eqn:f2LT-sum}
\end{align} 
which is similar to the BC sum rule for $g_2$
in Eq.\,(\ref{eqn:bc-sum}).
We note that the local operators in Eq.\,(\ref{eqn:operator-R})
are defined for $n \ge 2$, so that 
the WW relation and BC sum rule could not be rigorously proven 
in the operator-product-expansion formalism 
\cite{br-book,Blumlein:1996vs,Anselmino:1994gn,Deur:2018roz};
however, the WW relation and the BC sum rule are consistent 
with other derivations. 
Because the WW- and BC-like relations exist for both $f^+$ and $f^-$,
the same relations are valid for $f$ and $\bar f$ separately.

In this way, we confirmed the WW- and BC-like relations derived in
Ref.\,\cite{Kumano:2021fem} by the independent formal method.
Therefore, these relations can be used as reliable first estimate
of the twist-3 function $f_{LT} (x)$ in future theoretical and
experimental studies.
The JLab experimental results will be reported in the near future
on the tensor-polarized deuteron. The JLab data could contain 
significant higher-twist contributions in a few GeV$^2$ region, 
so that this work should be valuable.
There are recent theoretical estimates of $f_{LT}$ for the deuteron
\cite{Kumano:2025rai} and the $\rho$ meson \cite{Puhan:2025ujg}.
The $f_{LT}$ of the deuteron could be measured
in semi-inclusive DIS with the tensor-polarized deuteron target
\cite{Zhao:2025vol} and proton-deuteron Drell-Yan process \cite{Qiao:2024bgg}
by looking at the dependence of the azimuthal-angle ($\phi_{LT}$)
of the transverse tensor-polarization vector $S_{LT}^{\, \mu}$.

The present work is limited to the quark distributions. 
Similar relations have not been investigated for gluon distributions
because higher-twist tensor-polarized gluon distributions
are not formulated yet for spin-1 hadrons, although
analogous relations were derived for spin-1/2 hadrons
\cite{Hatta:2012jm,Koike:2019zxc}.

The tensor-polarized PDFs contain different aspects of high-energy
spin physics from the ones of the spin-1/2 nucleon. 
Especially, they are sensitive to dynamical aspects beyond
the simple addition of proton and neutron structure functions.
A new spin-physics field could be created by investigating
spin-1 PDFs and structure functions along the experimental projects 
at the aforementioned lepton and hadron accelerator facilities
starting from the JLab experiment.

\section{Summary}
\label{summary}

Using the OPE method with the local operators,
we derived the WW-like relation and the BC-like sum rule
for the twist-3 tensor-polarized PDF $f_{LT}$.
The results confirm the relations obtained in Ref.\,\cite{Kumano:2021fem},
where the matrix element of the non-local operator was expanded 
in terms of possible Lorentz vectors,
by an independent formal method. Now, these relations can be used
as reliable ones in future theoretical and experimental studies.
In particular, the JLab data will be reported 
for the tensor-polarized deuteron and they could contain
higher-twist effects, so that our studies should be valuable
in analyzing the experimental data and in making theoretical models.

\section*{Acknowledgments}

The authors were supported by the Chinese Academy of Sciences.
KK was also supported by the Gansu-province postdoctoral foundation.
They thank Qin-Tao Song for suggestions.



\begin{thebibliography}{00}
\bibitem{Kumano:2024fpr}
    S. Kumano, 
    \href{https://doi.org/10.1140/epja/s10050-024-01411-6}
    {Euro. Phys. J. A 60 (2024) 205}
    and references therein.
\bibitem{Fu:2026mb}
    D.-Y. Fu, Y.-B. Dong, S. Kumano, and J.-J. Xie, 
      \href{https://doi.org/10.1103/qq1h-snhk}
      {Phys.Rev.D 113 (2026) L111901}
    and references therein.
\bibitem{jlab-b1-2023} 
     K. Allada {\it et al}.,
     \href{https://www.jlab.org/exp_prog/proposals/13/PR12-13-011.pdf}
     {Proposal to Jefferson Lab PAC-40, PR12-13-011 
            (Update to PR12-11-110) (2023)}.
\bibitem{Poudel:2025nof}
    J. Poudel, A. Bacchetta, J.-P. Chen, and N. Santiesteban, 
    \href{https://doi.org/10.1140/epja/s10050-025-01558-w}
    {Euro. Phys. J. A 61 (2025) 81}.
\bibitem{Fermilab-SeaQuest-2017}
    M. Brooks {\it et al}., 
    \href{https://twist.phys.virginia.edu/work/E1039proposal_final.pdf}
    {Fermilab proposal (2017)}.
\bibitem{Keller:2020wan}
    D. Keller, D. Crabb, and D. Day, 
    \href{https://doi.org/10.1016/j.nima.2020.164504}
    {Nucl. Instrum. Meth. A 981 (2020) 164504}.
\bibitem{Arbuzov:2020cqg}
    A. Arbuzov {\it et al}.,
    \href{https://doi.org/10.1016/j.ppnp.2021.103858}
    {Prog. Nucl. Part. Phys. 119 (2021) 103858}.
\bibitem{Aidala:2019pit}
     C. A. Aidala {\it et al}.,
     \href{https://arxiv.org/abs/1901.08002}{arXiv:1901.08002}.
\bibitem{AbdulKhalek:2021gbh}
    R. Abdul Khalek {\it et al}.,
    \href{https://doi.org/10.1016/j.nuclphysa.2022.122447}
    {Nucl. Phys. A 1026 (2022) 122447}.
\bibitem{Anderle:2021wcy}
    D. P. Anderle {\it et al}.,
    \href{https://doi.org/10.1007/s11467-021-1062-0}
    {Front. Phys. 16 (2021) 64701}.
\bibitem{Bacchetta:2000jk}
    A.~Bacchetta and P.~Mulders 
    \href{https://doi.org/10.1103/PhysRevD.62.114004} 
    {Phys. Rev. D 62 (2000) 114004}.
\bibitem{Kumano:2020ijt}
    S.~Kumano and Q.-T.~Song,
    \href{https://doi.org/10.1103/PhysRevD.103.014025}
    {Phys. Rev. D 103 (2021) 014025}.
\bibitem{Kumano:2021fem}
    S. Kumano and  Q.-T.~Song,
    \href{https://doi.org/10.1007/JHEP09(2021)141}
    {J. High Energy Phys. 2021 (2021) 141}.
\bibitem{Wandzura:1977qf}
   S. Wandzura and F. Wilczek,
    \href{https://doi.org/10.1016/0370-2693(77)90700-6}
    {Phys. Lett. B 72 (1977) 195}.
\bibitem{Burkhardt:1970ti}
    H. Burkhardt and W. N. Cottingham,
    \href{https://doi.org/10.1016/0003-4916(70)90025-4}
    {Ann. Phys. 56 (1970) 453}.
\bibitem{Kumano:2025rai}
    S. Kumano and K. Kuroki,
    \href{https://doi.org/10.1016/j.physletb.2026.140348}
    {Phys. Lett. B 875 (2026) 140348}.
\bibitem{Jaffe:1989xx}
    R.~L.~Jaffe,
        {Comm. Nucl. Part. Phys. 19 (1990) 239}.
\bibitem{Jaffe:1996zw}    
    R.~L.~Jaffe,
        \href{https://arxiv.org/abs/hep-ph/9602236}   
        {arXiv:hep-ph/9602236}.
\bibitem{Anikin:2001ge}
    I.~V.~Anikin and O.~V.~Teryaev, 
        \href{https://doi.org/10.1016/S0370-2693(01)00555-X}
        {Phys. Lett. B 509 (2001) 95}.
\bibitem{Kodaira:1998jn}       
    J. Kodaira and K. Tanaka,
    \href{https://doi.org/10.1143/PTP.101.191}   
    {Prog. Theor. Phys. 101 (1999) 191}.
\bibitem{Kanazawa:2015ajw}
    K. Kanazawa, Y. Koike,  A. Metz1, D. Pitonyak, and M. Schlegel,
     \href{https://doi.org/10.1103/PhysRevD.93.054024}
     {Phys. Rev. D 93 (2016) 054024}.
\bibitem{Song:2023ooi}
    Q.-T. Song, 
     \href{https://doi.org/10.1103/PhysRevD.108.094041}
        {Phys. Rev. D 108 (2023) 094041}.
\bibitem{br-book} 
    V. Barone and R. G. Ratcliffe, 
       \href{https://doi.org/10.1142/5052}
       {Transverse Spin Physics (World Scientific, Singapore, 2003)}.
\bibitem{Blumlein:1996vs}
    J. Bl\"umlein and N. Kochelev, 
       \href{https://doi.org/10.1016/S0550-3213(97)00234-4}
       {Nucl. Phys. B 498 (1997) 285}.
\bibitem{Anselmino:1994gn}
    M. Anselmino, A. Efremov, and E. Leader,
    \href{https://doi.org/10.1016/0370-1573(95)00011-5}
    {Phys. Rept. 261 (1995) 1}.
\bibitem{Deur:2018roz}
   A. Deur, S. J. Brodsky, and G. F de Teramond, 
    {Rep. Prog. Phys. 82 (2019) 076201}.
\bibitem{Puhan:2025ujg}
    S. Puhan, S. Sharma, N. Kumar, and H. Dahiya,
    \href{https://doi.org/10.1103/xjkc-l5tr}   
    {Phys. Rev. D 113 (2026) 036030}.
\bibitem{Zhao:2025vol} 
    J. Zhao, A. Bacchetta, S. Kumano, T. Liu, and Y.-J. Zhou,
    \href{https://doi.org/10.1007/JHEP12(2025)067}   
    {JHEP {\bf 12} (2025) 067}.
\bibitem{Qiao:2024bgg}
    S.-Y. Qiao and Q.-T. Song, 
    \href{https://doi.org/10.1103/PhysRevD.111.054026}   
    {Phys. Rev. D {\bf 111} (2025) 054026}.
\bibitem{Hatta:2012jm}
    Y. Hatta, K. Tanaka, and S. Yoshida,
    \href{https://doi.org/10.1007/JHEP02(2013)003}   
    {JHEP {\bf 02} (2013) 003}.
\bibitem{Koike:2019zxc}
    Y. Koike, K. Yabe, and S. Yoshida, 
    \href{https://doi.org/10.1103/PhysRevD.101.054017}   
    {Phys. Rev. D {\bf 101} (2020) 054017}.
\end{thebibliography}
\end{document}